\documentclass[superscriptaddress,reprint,
 amsmath,amssymb,
 aps,
]{revtex4-2}
\usepackage[utf8]{inputenc}
\usepackage{multirow}
\usepackage{mathtools}
\usepackage{caption}
\usepackage{babel}
\usepackage{graphicx}% Include figure files
\usepackage{dcolumn}% Align table columns on decimal point
\usepackage{bm}% bold math
\usepackage{cleveref}
\usepackage{xcolor}
\usepackage{amsmath}
\usepackage{physics}

\setcitestyle{super}

\newcommand*{\citen}[1]{%
  \begingroup
    \romannumeral-`\x % remove space at the beginning of \setcitestyle
    \setcitestyle{numbers}%
    \cite{#1}%
  \endgroup   
}

\newcommand{\gowo}{G_\text{0}W_\text{0}}

\begin{document}

\author{Vibin Abraham}
\email{avibin@umich.edu}
\affiliation{%
 Department of Chemistry, University of Michigan, Ann Arbor, MI 48109.
}%

\author{Gaurav Harsha}
\affiliation{%
 Department of Chemistry, University of Michigan, Ann Arbor, MI 48109.
}%
\author{Dominika Zgid}
%\email{zgid@umich.edu}
\affiliation{%
 Department of Chemistry, University of Michigan, Ann Arbor, MI 48109.
}%
\affiliation{%
 Department of Physics and Astronomy, University of Michigan, Ann Arbor, MI 48105.
}%

\date{\today}

\title{Relativistic fully self-consistent $GW$ for molecules: Total energies and ionization potentials}
\begin{abstract}

The fully self-consistent $GW$ (sc$GW$) method with the iterative solution of Dyson equation provides a consistent approach for describing the ground and excited states without any dependence on the mean-field reference.
    In this work, we present a relativistic version of sc$GW$ for molecules containing heavy element using the exact two-component (X2C) Coulomb approximation.
We benchmark \texttt{SOC-81} dataset containing closed shell heavy elements for the first ionization potential using the fully self-consistent $GW$ as well as one-shot $GW$.
The self-consistent $GW$ provides superior result compared to $\gowo$ with PBE reference and comparable to $\gowo$ with PBE0 while also removing the starting point dependence.
The photoelectron spectra obtained at the X2C level demonstrate very good agreement with experimental spectra.
We also observe that sc$GW$ provides very good estimation of ionization potential for the inner $d$ shell orbitals. 
Additionally, using the well conserved total energy, we investigate the equilibrium bond length and harmonic frequencies 
    of few halogen dimers using sc$GW$. 
Overall, our findings demonstrate the applicability of the fully self-consistent $GW$ method for accurate ionization potential, photoelectron spectra and total energies in finite systems with heavy elements with a reasonable computational scaling.
\end{abstract}
\maketitle

\section{Introduction}
The $GW$ method~\cite{Hedin1965,Onida2002,Golze2019,Reining2018,Brunevalgwreview} is a powerful approach introduced in solid-state physics to calculate electronic properties of materials as it can accurately predict fundamental properties such as band gaps, quasiparticle energies, ionization potentials (IP) and electron affinities (EA).
These properties are crucial for understanding and designing materials with unique functionalities. 
In the recent years, the $GW$ approach has also been applied to molecules to study valence photoelectron spectra, IPs and EAs\cite{Stan2009,Rostgaard2010,Foerster2011,Bruneval2013,Vlcek2017,Korbel2014,Ren2012,van2015,Krause2015,Forster2021,Caruso2016,Mejia2021,Duchemin2020,kaplan2016quasi} and X-ray photoelectron spectra (XPS).\cite{van2018,Golze2020,Galleni2022,Li2022,Mejia2022}

The first implementations of the $GW$ equations by Hybertsen et al.~\cite{Hybertsen1985} and Godby et al.~\cite{Godby1986} performed only one iteration of the $GW$ self-consistency cycle, in addition to invoking other approximations.
It is, therefore, called as the one-shot $GW$ or the $\gowo$ approximation, and has been applied to solids\cite{Gonze2005,Shishkin2006,west2022,DESLIPPE2012,Zhu2021} and molecules.\cite{van2013,Bruneval2013,van2015,Forster2021,Vlcek2017,marie2023gw,Wang2022,Ke2011}
As computational capabilities have come a long way since the 1980s, full implementation of the original $GW$ theory, also known as the self-consistent $GW$ (sc$GW$), have become possible for realistic systems.~\cite{Stan2009,Rostgaard2010,Caruso2012,Yeh2022,Grumet2018,Strange2011,Ren2012,Caruso2013,Yeh2022,wen2023comparing}

For molecules, sc$GW$ has exhibited significant promise as it provides accurate results for ionization potential, dipole moments, ground state density, and total energies, often closely matching CCSD results.\cite{Caruso2012,Caruso2013,Caruso2014,wen2023comparing}
The sc$GW$ method also provides an unbiased estimation of correlation energy which, due to self-consistency, is independent of the method with which it is calculated. 
It also offers unbiased assessments of both thermodynamic and spectral properties independent of the initial starting point.

The relatively low computational cost ($\mathcal{O}(n^4)$) compared to other wavefunction methods makes $GW$ an attractive alternative, especially when the system is large.
As one goes to heavier elements, relativistic effects also become significant.\cite{pyykko2012}
% The $GW$ method proves particularly well-suited for such systems due to its relatively low computational cost alongside its high accuracy.
Relativistic generalization of $GW$ theory proves particularly well-suited for such systems.
Consequently, relativistic corrections, including spin-orbit coupling (SOC) effects, have become more prevalent in $GW$ over the past decade, applied to both solids\cite{Sakuma2011,Kutepov2012,Brivio2014,Ahmed2014,Molina2013,Irene2013,Irene2015,umari2014relativistic} and molecules.\cite{Kuhn2015,Korbel2014,Scherpelz2016,Holzer2019,Franzke2022,Forster2023,Franzke2023} 
While Refs.~\citen{Kutepov2012,Yeh2022x2c} have examined relativistic corrections in fully self-consistent $GW$ for solids, there's a lack of systematic exploration regarding sc$GW$ with relativistic corrections for molecules.
This paper aims to remedy this issue by investigating relativistic sc$GW$ for molecular problems.

Even for the one-shot $GW$, challenges in obtaining consistent results persist when comparing different $GW$ results, primarily stemming from differences in implementation details such as the choice of single-particle basis sets, real- or imaginary-time and frequency grid configurations, analytic continuation method, etc.
Recently, the $GW$100 molecular dataset\cite{van2015} was proposed to compare different implementations.\cite{Govoni2018,Maggio2017,Caruso2016,Forster2021}
Additional complications arise when systems with heavy elements are investigated.
Along these directions, Scherpelz\cite{Scherpelz2016} and coworkers made a benchmark of 81 molecules (\texttt{SOC-81}) and studied the $\gowo$ approximation using the WEST code.\cite{Govoni2015,west2022}
Later, F\"{o}rster et al.~\cite{Forster2023} also studied the \texttt{SOC-81} benchmark with their two-component $GW$ implementation with Slater type orbitals (STO) with vertex-corrected $GW$ methods and quasiparticle self-consistent $GW$.
Even though Gaussian orbitals are frequently used in molecular calculations, their use in relativistic $GW$, is scarce and, to our knowledge, a rigorous benchmark study for relativistic sc$GW$  using
Gaussian orbitals for molecules has not been performed so far.

In this work, we present a detailed analysis of $GW$ methods for molecules containing heavy elements, by using Gaussian orbitals.
In particular, we examine different properties, namely ionization potentials, total energies, and spin-orbit splittings in inner-valence photoelectron spectra to establish the competency of the self-consistent $GW$.
We also look at the crystal-field and spin-orbit splittings in outermost $d$-orbitals in selected transition metal compounds.
All the tests are performed using the \texttt{SOC-81} dataset of relativistic molecules.
In order to include SOC effects, we employ a two-component framework\cite{van1993,van1996,van1996a,Nakajima2012,Reiher2012,Liu2007,Dyall1997,Dyall2001} 
    using the exact two-component (X2C) formalism.~\cite{Liu2009,Kutzelnigg2005,Liu2007,Sun2009,Cheng2011}

\section{Theory}
In this section, we first describe the relativistic formalism and the resulting X2C Hamiltonian.
We then discuss the $GW$ approximation, with special focus on the one-shot and self-consistent methods.
While a detailed discussion of these topics is beyond the scope of this work, we here present information necessary for a self-contained discussion.
% In this section, we briefly discuss the X2C Hamiltonian and the different levels of $GW$ method.

\subsection{X2C Hamiltonian}
The X2C formalism has been widely explored in previous studies.\cite{Liu2009,Cheng2011,Liu2018,Asthana2019}
In this context, we provide a concise overview of the fundamental principles.
The X2C Hamiltonian can be derived from the Dirac Hamiltonian using kinetically balanced basis set.\cite{Dyall2001}
%The exact two-component Hamiltonian is generated by a two step transformation.\tocite{Foldy}
The non interacting four-component Dirac Hamiltonian in the matrix form can be written as:

\begin{equation}
\mathbf{H}^{\text{4C1e}} = 
\left[\begin{array}{cc}
\hat{V} & \hat{T} \\
\hat{T} & \frac{1}{4 c^2} \hat{L}-\hat{T},
\end{array}\right]
\end{equation}
where $\hat{V}$ is the local Coulomb potential,  $\hat{T}$ is the kinetic energy matrix, 
%$\hat{S}$ is the overlap matrix, 
and $c$ is the speed of light. 
The $\hat{L}$ operator is:
\begin{equation}
    \hat{L} = \bra{\Phi}  (
        \vec{\sigma} \cdot \vec{p} ) \hat{V} (\vec{\sigma} \cdot \vec{p}
    ) \ket{\Phi},
\end{equation}
where $\vec{\sigma}$ is the vector of Pauli spin matrices and $\vec{p}$ is the momentum operator.
Using Dirac identity, the $\hat{L}$ matrix can be expressed in terms of spin-dependent and spin-free contributions, defined as
\begin{equation}
    \hat{L} = \bra{\Phi} \vec{p} \cdot (\hat{V} \vec{p}) \ket{\Phi} + \bra{\Phi} i \vec{\sigma}  \cdot (\vec{p} \times \hat{V} \vec{p}) \ket{\Phi}.
\end{equation}
Approximating $\hat{L}$ with the first (spin-independent) term alone leads to the spin-free X2C-1e (sfX2C-1e) approximation.\cite{Cheng2011}
Inclusion of both the spin-dependent and spin-free terms constitute the X2C-1e approximation.
In the discussion below, we use X2C-1e and X2C acronyms interchangeably to indicate the two-component approximation, and similarly, we use sfX2C or sfX2C-1e for scalar relativistic approximation. A particular level is obtained, as discussed earlier, by approximations performed on $\hat{L}$.

Within X2C, the exact coupling between the large and small components, is obtained by diagonalizing the one-body Dirac-Coulomb Hamiltonian:
\begin{equation} 
    U^{\dagger} \mathbf{H}^{\text{4C1e}}  U=  \left[\begin{array}{cc}
        \mathbf{H}_{+}^{\text{X2C1e}} & 0 \\
        0 & \mathbf{H}_{-}^{\text{X2C1e}}
    \end{array}\right].
    \label{eq:x2c-ham}
\end{equation}
Here, $\mathbf{H}_{+}^{\text{X2C1e}}$ Hamiltonian exactly reproduces the eigenvalues of the electronic degrees of freedom of the parent four-component Hamiltonian.
In the final $\mathbf{H}_{+}^{\text{X2C1e}}$ Hamiltonian, spin dependence term only comes from $\hat{L}$.\cite{Dyall1994,Dyall2001}
This one-electron Hamiltonian is then used with the non-relativistic two-electron integrals. 
The SOC effects in the two-electron integrals, which are neglected in the X2C-1e formalism, are mainly important for fine structure splitting and inner core orbital properties.
This approximation leads to the `picture-change' error since same transformation for two-electron integral is not done as for one-electron (or 1e) Hamiltonian and approximate large component integrals are used instead.
The non-relativistic two-electron integrals are given as
\begin{equation}
    (pq|rs)=\int d r_1 d r_2 \phi_{p}(r_1)^* \phi_{q}(r_1) \frac{1}{r_{12}} \phi_{r}(r_2)^* \phi_{s}(r_2),
\end{equation}
where $\phi_{p}(r)$ denotes the one-particle wave function corresponding to the $p$th atomic orbital.
In our implementation of the $GW$ method, we use the density fitting approximation which significantly decreases the computational cost.\cite{Ren2012,Werner2003,Yeh2022} 
The two-electron integrals are then written as product of two three-index tensors.
\begin{equation}
    (pq|rs) = \sum_Q V_{pq}^{Q} V_{rs}^{Q},
    \label{eq:df-approx}
\end{equation}
where $p$, $q$, $r$, and $s$ are atomic orbital (AO) indices and $Q$ is an auxiliary basis index.

\subsection{$GW$ approximation}

The $GW$ approximation to the self-energy is the simplest non-trivial approximation to Hedin's perturbation theory, where the self-energy is defined as the product of the one-particle Green's function $G$ and the screened Coulomb interaction $W$.
\color{black}
For all the results presented in this paper, we use the $GW$ method as implemented in Refs.~\citen{Yeh2022,Yeh2022x2c,Iskakov2020,Pokhilko2021}
While the interested reader can find a detailed description of the implementation in these works, here, for clarity we only briefly discuss the main quantities involved.

The finite temperature, Matsubara Green's function in atomic orbitals is defined as:
\begin{align}
    \mathbf{G}\left(i \omega\right)
    &= \left[\left(i \omega+\mu\right) \mathbf{S}-\left(\mathbf{H}_{+}^{\mathrm{X} 2 \mathrm{C} 1 \mathrm{e}}\right)-\mathbf{\Sigma}\left(i \omega\right)\right]^{-1},
    \label{eq:dyson}
\end{align}
where $i \omega$ is the Matsubara frequency, $\mu$ is the chemical potential, $S$ the overlap matrix, and $\Sigma (i\omega)$ is the self-energy.
The one-body Hamiltonian matrix $\textbf{H}_+^{\text{X2C1e}}$, defined in \cref{eq:x2c-ham}, contains the kinetic term, the external potential (e.g., from background ions), as well as the terms arising from the X2C (or sfX2C) approximation.

The $GW$ self-energy can be separated into $\omega$-independent (static) and $\omega$-dependent (dynamical) contributions,
\begin{equation}
    \mathbf{\Sigma} (i \omega) = \mathbf{\Sigma}^\infty + \mathbf{\Sigma}^c (i \omega),
\end{equation}
where $\Sigma^\infty$ contains the Hartree and exchange terms, while $\Sigma^c$ contains contributions from the RPA bubble diagram.
In other words,
\begin{subequations}
    \begin{align}
        \Sigma^{\infty}_{p\sigma, q\sigma'} &=  \sum_{rs} [V_{pq}^{Q}V_{rs}^{Q}\rho_{r\sigma, s\sigma'} + V_{ps}^{Q}V_{rq}^{Q}\rho_{r\sigma, s\sigma'} ] ,
        \label{eq:sigma_static}
        \\
        \Sigma^{c}_{p \sigma, q \sigma'} (\tau) &=
        \sum_{r s} \sum_{Q Q^{\prime}} G_{r \sigma, s \sigma'}(\tau) V_{pr}^Q \tilde{P}_{Q Q^{\prime}}(\tau) V_{qs}^{Q^{\prime}},
    \end{align}
\end{subequations}
where $\tilde{P}$ is the polarization function, $\beta$ corresponds to the inverse temperature used in the calculation, $\tau \in [0, \beta]$ is the imaginary time and $\rho$ is the one-particle density matrix, which is related to the imaginary-time Green's function as
\begin{equation}
    \mathbf{\rho} = -G (\tau = \beta^-).
\end{equation}
To compute $\tilde{P}$, one first obtains the irreducible or the non-interacting polarization $\tilde{P}_0$, the computation of which is generally very expensive since it results in a frequency-dependent four-index tensor.
This computational cost is brought down by using the density-fitting approximation and constructing the polarization function in the auxiliary basis.
This step also avoids the explicit formation of the screened interaction ($W$), and is often done in modern $GW$ codes.\cite{Caruso2012,Yeh2022,Ren2012,Forster2021,Duchemin2021}
On the imaginary-time axis, ${P}_0$ is simply the product of $G(-\tau)$ and $G(\tau)$,
\begin{equation}
    \tilde{P}_{0}^ {Q Q^{\prime}}(\tau) 
    = -\sum_{\sigma \sigma^{\prime}} \sum_{a b c d} V_{d a}^Q 
    G_{c \sigma^{\prime}, d \sigma}(-\tau)  G_{a \sigma, b \sigma^{\prime}}(\tau) V_{bc}^{Q^{\prime}},
\end{equation}
where $Q$ and $Q'$ are auxiliary basis indices introduced in the density fitting approximation in \cref{eq:df-approx}.
The full polarization $\tilde{P}$ can then be obtained as
\begin{equation}
    \tilde{\mathbf{P}} (i\Omega) = \sum_{m=1}^\infty \left[ 
        \tilde{\mathbf{P}}_0 (i \Omega)
    \right]^m
    = \left[
        \mathbb{I} - \tilde{\mathbf{P}}_0 (i \Omega)
    \right]^{-1} \tilde{\mathbf{P}}_0 (i\Omega),
\end{equation}
where $\mathbb{I}$ is the identity matrix of dimension equal to the number of auxiliary basis functions, and $\Omega$ is the bosonic Matsubara frequency. 
Note that the spin-orbit effects in $\tilde{P_0}$ and $\tilde{P}$ appear via summed-over spin degrees-of-freedom in the Green's function.

The starting point of $GW$ calculations is a density functional theory (DFT) or Hartree-Fock (HF) solution, in which case $\Sigma(i\omega)$ is initialized as the effective Hartree and exchange potential from the mean-field method.
The chemical potential $\mu$ is optimized to ensure the correct number of electrons.
The two-component formalism is solved using the complex Generalized Hartree-Fock (GHF) method to include the X2C Hamiltonian.\cite{Desmarais2019,Kadek2019,Sun2019ghf,goings2015stability}
The GHF method enables the self-consistent treatment of electron-electron interactions as well as a non-perturbative inclusion of spin-orbit effects, making it suitable for describing complex electronic structure.
%mention comp cost? 

\subsubsection{One-shot $GW$ ($\gowo$)}

As the name suggests, in one-shot $GW$ or $\gowo$, only one iteration of the $GW$ equations is carried out.
The one-shot $GW$ is often used in conjunction with the diagonal approximation, which is based on the assumption that the mean-field eigenstates are a good approximation to the $GW$ eigenstates, even though the eigenvalues are not.
In the diagonal approximation, the off-diagonal elements of $\Sigma$ (in the basis of mean-field eigenstates) are discarded.
This results in a non-linear relation between the mean-field and $\gowo$ quasiparticle energies, denoted, respectively, by $\epsilon^0$ and $\epsilon$,
\begin{equation}
    \epsilon_i = \epsilon_i^0 + \bra{\Phi_i} \text{Re} \Sigma (\epsilon_i) - v^{xc}  \ket{\Phi_i},
    \label{eq:quasiparticle}
\end{equation}
where $v^{xc}$ is the exchange correlation potential corresponding to the reference mean-field method.
In practice, we start with the mean-field (typically DFT or HF) Green's function, and then construct the self-energy matrix on the Matsubara frequency axis, $\Sigma (i \omega)$.
This is followed with analytic continuation of $\Sigma$ from Matsubara to real-frequency axis, where the quasiparticle equation, or \cref{eq:quasiparticle}, is solved.
We have employed the Nevanlinna approach~\cite{Fei2021} to perform the analytic continuation.

\subsubsection{Self-consistent $GW$ (sc$GW$)}
The quasiparticle equation can be understood as a perturbative correction on top of the mean-field quasiparticle energies.
As is apparent from \cref{eq:quasiparticle}, one-shot $GW$ has a strong dependence on the initial mean-field solution, whether it be Hartree-Fock or different functionals used in DFT.
Self-consistency of the $GW$ equations is necessary to alleviate this dependence. 

\begin{figure}
    \centering
    \includegraphics[width=\linewidth]{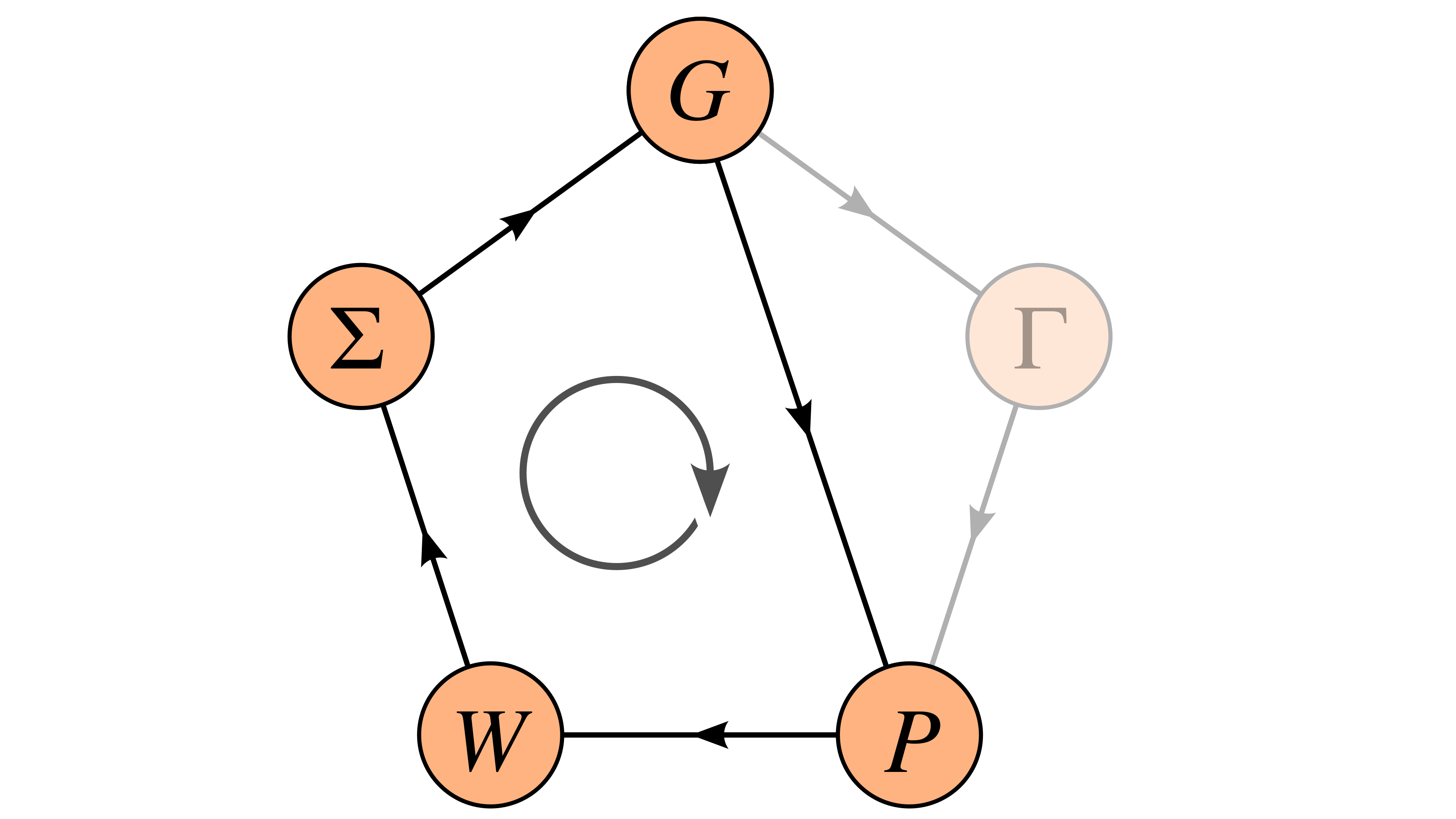}
    \caption{Schematic of Hedin's pentagon. In the scGW algorithm all quantities except for $\Gamma$ is solved self consistently.}
    \label{fig:hedin}
\end{figure}

Our algorithm for sc$GW$ is based on iterative solution of the Dyson equation, or \cref{eq:dyson}.
The Hedin's equations are solved iteratively excluding the vertex function ($\Gamma$) as shown in Figure \ref{fig:hedin}.
The Green's function from the previous iteration is used to compute the polarization matrix, which is then used to obtain the dynamical part of self-energy.
Similarly, the density matrix $\rho$ coming from the same Green's function is used to compute the static self-energy.
Plugging the new self-energy in \cref{eq:dyson}, and adjusting the chemical potential $\mu$ to fix the number of electrons, we obtain the new Green's function.
This is repeated until the total energy as well as the energy from the static self-energy alone are converged to a very tight threshold of $10^{-6} E_h$.

\subsubsection{sc$GW$: Total Energies}
At self-consistency, the thermodynamic properties are well-defined, meaning that different ways to evaluate the energy and other thermodynamic properties in sc$GW$ lead to the same results.~\cite{Baym1961, Baym1962, Holm1999, Dahlen2006, Stan2009}
In this work, we use the Galitskii–Migdal formula~\cite{Holm2000} to compute the correlation energy, which leads to the following expression for the total energy:
\begin{subequations}
    \begin{align}
        E_\mathrm{Total} &= E_\mathrm{nuclear} + E_{1b} + E_{2b},
        \\
        E_{1b} &= \mathrm{Tr} \left[
            \rho \left(
                \mathbf{H}^{X2C1e}_+ + \mathbf{\Sigma}^{\infty}
            \right)
        \right],
        \\
        E_{2b} &= \sum_{n = -\infty}^{\infty} \mathrm{Tr} \left[
            \mathbf{G} (i\omega_n) \mathbf{\Sigma}^c (i \omega_n)
        \right].
    \end{align}
\end{subequations}
Here, $E_\mathrm{nuclear}$ denotes the electrostatic energy due to repulsion between the fixed atomic nuclei.
The Galitski-Migdal formula and the derivative of the grand-potential, both provide identical estimates for the energy.\cite{Baym1962,Iskakov2019,Yeh2022}
%The self-energy expressions derived as functional derivatives of a $\Phi$ functional of the Green function possess a crucial property: they yield many-body approximations that conserve fundamental quantities such as particle number, momentum, angular momentum, and energy. 

\section{Computational details}\label{sec:comp-details}
The reference DFT or HF calculations and the density-fitted integrals are obtained from \texttt{PySCF} version 2.2.1. \cite{Sun2020}
We use the \texttt{X2C-TZVPall} and \texttt{X2C-QZVPall} basis sets\cite{pollak2017a} for all the calculations reporting ionization energies.
The IPs obtained in the triple- and quadruple-$\zeta$ calculations are then extrapolated to the complete basis-set limit, assuming an inverse relationship with the number of atomic orbital basis functions.~\cite{van2015, Forster2021}
For the inner ionization peaks and $d$-orbital IPs in Section \ref{subsec:inner-ip}, as well as for total energies in Sec.~\ref{subsec:total-energy}, we directly use the data obtained using \texttt{X2C-QZVPall} basis set, without any extrapolation.
For all calculations, even-tempered Gaussian orbitals~\cite{Stoychev2017} with progression factor of $2.0$ were used to perform the density fitting for the two-electron integrals in \texttt{PySCF}.\cite{Sun2020}

All dynamic quantities in $GW$ are expanded using the efficient intermediate representation grids, employing sparse sampling along both the imaginary-time and Matsubara frequency axes.\cite{Shinaoka2017,Li2020}
All $GW$ calculations are performed at $\beta =100$ a.u.${}^{-1}$.
For sc$GW$, we use the direct inversion of iterative subspace (DIIS) to accelerate convergence.~\cite{Pokhilko2022}
For all the calculations in this work, converging the total $GW$ energy to within $10^{-6} E_h$ required a maximum of 20 iterations.

For the self-consistent $GW$ calculations, the spectral function and consequently the ionization potential is obtained by analytical continuation of the converged single-particle Green's function from the imaginary to the real frequency axis using the Nevanlinna analytical continuation method.\cite{Fei2021}
For $\gowo$, similar analytic continuation was performed for the diagonal of the self-energy (in the appropriate mean-field molecular orbital basis) after first iteration of $GW$.
To avoid singularities, all analytic continuations have been performed with a broadening of $\eta = 0.004$ a.u.

\subsection{\texttt{SOC-81}}
In this work, we utilize the \texttt{SOC-81}~\cite{Scherpelz2016} dataset for validating our $GW$.
The \texttt{SOC-81} has 81 closed-shell molecules that contain heavy elements with significant scalar relativistic effects as well as an increasing contribution from the spin-orbit coupling (SOC) for molecules containing heavier elements.
The \texttt{SOC-81} geometries were obtained from Ref.~\citen{Scherpelz2016}.
Experimental references for the \texttt{SOC-81} are taken from the NIST webbook.\cite{NIST}

On detailed examination, we identified discrepancies in the NIST database for HgCl$_2$, KBr, and RbBr. In place of the first IP, the second IP (resulting from the spin-orbit splitting) was documented for these compounds.
We have updated the reference values for these three systems, which have been utilized to benchmark the results presented in Section \ref{subsec:soc81}.
We have also excluded a total of 7 systems from the benchmark analysis. Two of these systems, MgI$_2$ and AsF$_5$, were excluded due to the unavailability of reliable experimental references. Additionally, the following five largest systems from the benchmark — Al$_2$Br$_6$, (C$_5$H$_5$)$_2$Ru, Mo(CO)$_6$, TiI$_4$, ZrI$_4$ — were omitted as their computational requirements exceed our current capabilities.
%In case of MgI$_2$, the experimental values is reported same as MgCl$_2$ which is not correct and for AsF$_5$ a reliable reference is lacking in NIST webpage.
%It is experimentally challenging to obtain the IP for MgI$_2$ since it decays very fast.\cite{berkowitz1966,Aita1989}

\begin{figure*}
    \centering
    \includegraphics[width=\linewidth]{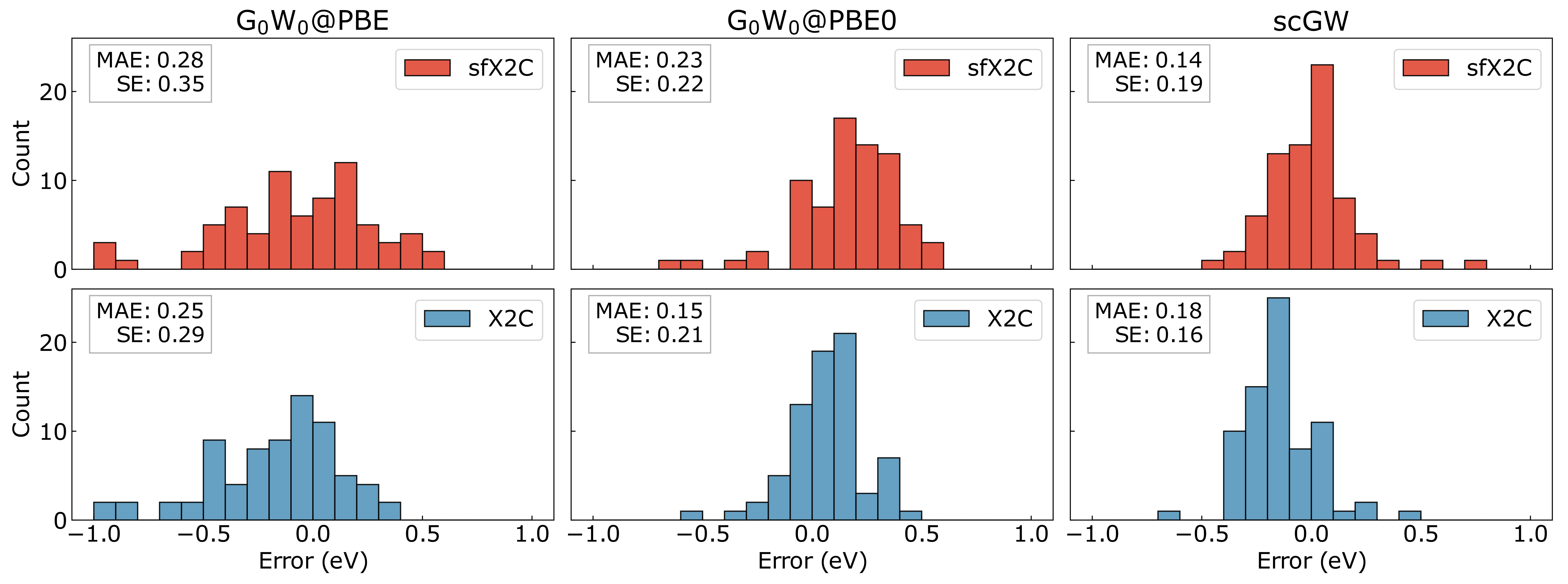}
    \caption{Histogram of errors with respect to experiment for ionization potential of $\gowo$ (with PBE and PBE0 reference) with the fully self-consistent $GW$ results with scalar relativistic correction (top) and two-component method (bottom). The errors are plotted in bins of 0.1 eV. The mean absolute error (MAE) and standard deviation of error (SE) is also reported for each method.}
    \label{fig:mad_$GW$}
\end{figure*}

\section{Results}

In this section, we discuss the results obtained using the $GW$ methods.
In section \ref{subsec:soc81}, we benchmark $G_0 W_0$ and sc$GW$ for the \texttt{SOC-81} database.
In section \ref{subsec:inner-ip}, we analyze the inner IP obtained using sc$GW$. 
In section \ref{subsec:total-energy}, we analyze the equilibrium bond length and harmonic frequency obtained from the total energies of sc$GW$.

\subsection{Vertical Ionization Potential for \texttt{SOC-81}}\label{subsec:soc81}

We benchmark our sfX2C- and X2C-$GW$ implementations by studying the vertical IPs for molecules in the \texttt{SOC-81}~\cite{Scherpelz2016} dataset using both sc$GW$ and $G_0 W_0$ with  Perdew-Burke-Ernzerhof~\cite{Perdew1996} (PBE) generalized gradient approximation (GGA) and hybrid PBE0~\cite{perdew1996rationale} (25\% exact exchange) references.

In \cref{fig:mad_$GW$}, we present the distribution of error in vertical IPs with respect to experiment, for $\gowo$@PBE, $\gowo$@PBE0 and sc$GW$ for the \texttt{SOC-81} benchmark.
We also present the statistical data in \cref{tbl:mae}.
At the scalar relativistic level (sfX2C), sc$GW$ performs the best among the three variants of $GW$ method, with a small mean absolute error (MAE) of 0.14 eV.
The standard deviation of error (SE) is also much smaller for sc$GW$.
For the one-shot methods, $\gowo$ with the hybrid PBE0 reference performs better than PBE for the \texttt{SOC-81} dataset.
An identical behavior for $\gowo$ has been observed previously in the plane-wave~\cite{Scherpelz2016} and STO~\cite{Forster2023} implementations. Furthermore, PBE0 is also considered as a better reference than PBE for $\gowo$ in non-relativistic calculations as well.~\cite{Bruneval2013,Zhang2022}

\begin{table}[]
    \caption{Comparison of mean absolute error (MAE), standard deviation of error (SE) and mean error (ME) of $G_0W_0@PBE$, $G_0W_0@PBE0$  and sc$GW$ with respect to experimental values. All values are reported in eV. }
\renewcommand{\arraystretch}{1.3}
\begin{tabular}{cccc|ccc}
\hline
\hline
\multicolumn{1}{l}{} & \multicolumn{3}{c|}{sfX2C} & \multicolumn{3}{c}{X2C} \\ \cline{2-7}
\multicolumn{1}{l}{Method} & MAE    & SE     & ME      & MAE    & SE    & ME     \\ \hline
$\gowo$@PBE             & 0.28   & 0.35   & -0.08   & 0.25   & 0.29  & -0.17  \\
$\gowo$@PBE0            & 0.23   & 0.22   & 0.16    & 0.15   & 0.21  & 0.05   \\
sc$GW$                 & 0.14   & 0.19   & -0.01   & 0.18   & 0.16  & -0.14  \\ \hline
\hline
\end{tabular}
    \label{tbl:mae}
\end{table}

For the fully relativistic corrections (X2C), both sc$GW$ and $\gowo$@PBE0 provide results with similar accuracy,
with MAE of 0.18 eV and 0.15 eV, respectively.
% $\gowo$@PBE0 has a smaller MAE of 0.15 eV whereas sc$GW$ has a MAE of 0.18 eV.
While $\gowo$@PBE0 has a marginally smaller MAE, sc$GW$ has a smaller SE of 0.16 eV compared to 0.21 eV in $\gowo$@PBE0.
The quality of $\gowo$ results, for PBE and PBE0 references, improves marginally at the two-component level compared to the scalar relativistic level of theory.

Generally, we see that sc$GW$ underestimates the IP peaks at the two-component level as compared to sfX2C as can be seen from the mean-error (ME).
This is probably because the two-electron contributions to the SOC corrections is not included.\cite{Neese2005,Sikkema2009,SCHIMMELPFENNIG1998}
Moreover, for several systems containing SOC splitting at the valence IP, the resultant double peaks are not clearly resolved in experiments.
This spectral complexity, as observed in CdCl$_2$ (discussed in section \ref{sec:cdx2}) and several others within the \texttt{SOC-81} dataset, impacts the accuracy of reported experimental IPs.
In such cases, the reference IPs are estimated using semi-empirical parametric models,~\cite{PRICE1974,ELAND1970,BANCROFT1983} whose accuracy is not controlled.
High resolution experimental results, backed by accurate quantum chemical methods, are desirable for further scrutinizing the accuracy of different $GW$ results and develop more accurate methods.

In the case of $\gowo$, the accuracy of results depend significantly on the selection of the DFT reference. Additionally, for a couple of systems, the quasiparticle equation exhibited multiple solutions, resulting in incorrect IP peak assignments.
This is discussed in detail in section \ref{subsec:qp}.
On the other hand, sc$GW$ results are independent of reference, making it a better predictive method.
Hence, in order to ensure consistent results, it is advisable to employ the self-consistent Green's function (sc$GW$) method, as it effectively eliminates the influence of the mean-field reference as well as the multiple solution issue.

\begin{figure*}
    \centering
    \includegraphics[width=\linewidth]{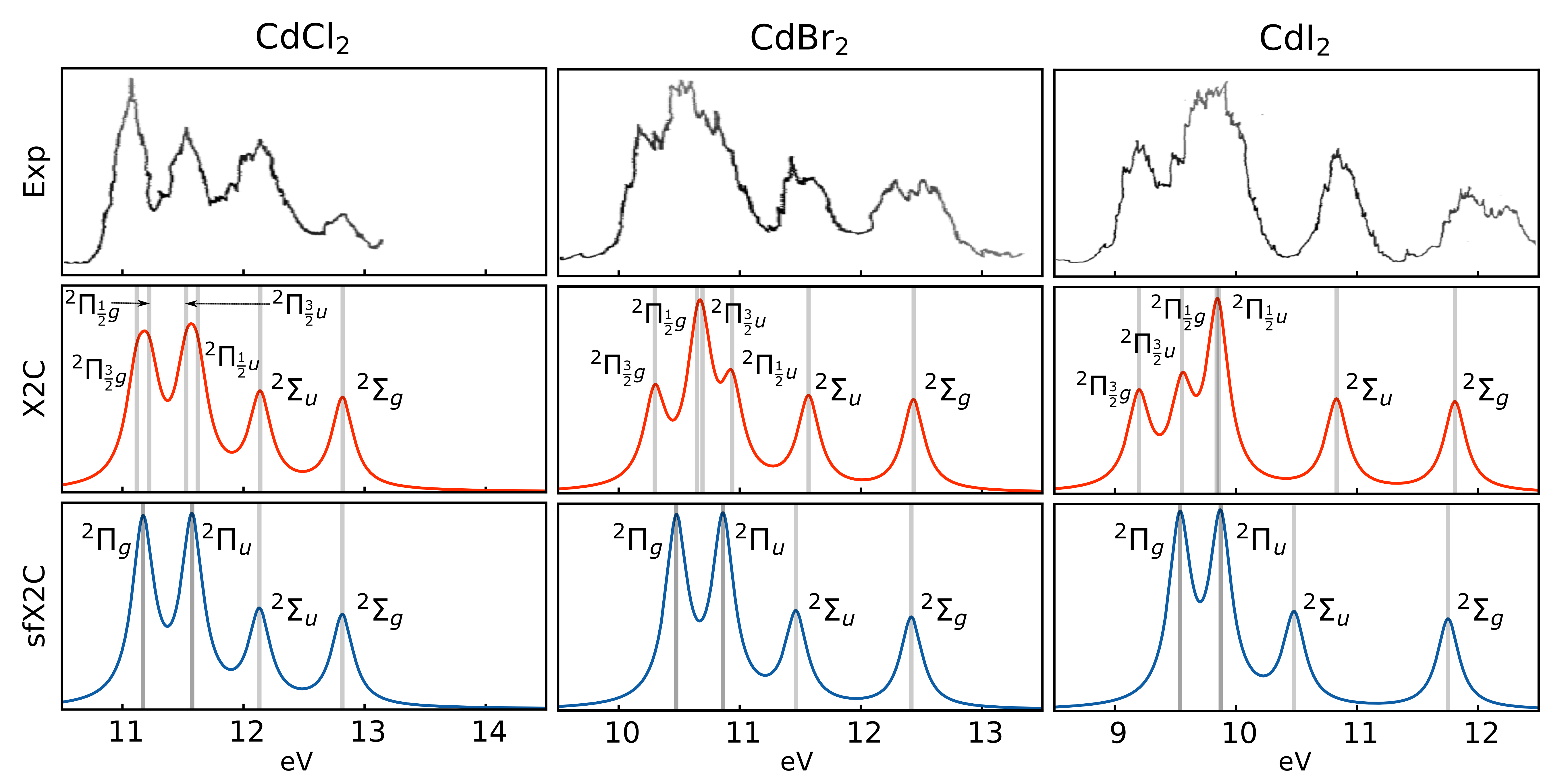}
    \caption{Comparison of experimental spectra with sc$GW$ spectra at the scalar relativistic and fully relativistic level for CdX$_2$ (X=Cl,Br,I).
    All three experimental spectra are taken from work by Boggess et al.\cite{BOGGESS1973} and shifted by 0.2 eV. The vertical lines in sfX2C and X2C represent the ionization energies. The SOC effect introduced in the X2C can be seen by tracking the vertical lines from sfX2C to X2C. Theoretical results are computed using the \texttt{X2C-QZVPall} basis set.}
    \label{fig:cdx}
\end{figure*}

\subsubsection{Multiple solution in quasiparticle equation in $\gowo$}\label{subsec:qp}
The non-linear quasiparticle (QP) equation, \cref{eq:quasiparticle}, in the $\gowo$ approach generally has multiple solutions.
Either of these solutions can be converged to, depending on the initial guess and numerical solver.~\cite{van2015,Scherpelz2016}
We find that the problem of converging to a wrong solution is prominent in $\gowo$@PBE for the \texttt{SOC-81} set, both at the level of sfX2C and X2C relativistic corrections.
The effect of these multiple solutions becomes less pronounced when a PBE0 reference is employed for $\gowo$.
Such problems are well known within the $\gowo$ literature.~\cite{van2015, Scherpelz2016}
Obtaining consistent quasiparticle solutions across different basis set orders is necessary for extrapolating results to the complete basis-set (CBS) limit.
Therefore, we also examine the $\gowo$ spectral function for obtaining the correct quasiparticle solution. The spectral function is defined as:
\begin{align}
    A_p (\omega)
    &= -\frac{1}{\pi} \mathrm{Im}[ G^0_{pp} (\omega)] \nonumber,
    \\
    &= -\frac{1}{\pi} \mathrm{Im} [\omega + i \eta - \epsilon^0_p - (\Sigma(\omega)_p + \Sigma^x_p - v^{xc}_p )] ^{-1},
\end{align}
where $\eta$ is a small broadening parameter, that is required to avoid poles on the frequency axis. As mentioned in Section~\ref{sec:comp-details}, we use $\eta = 0.004$ a.u.
By default, sc$GW$ avoids this issue because the IP is identified from the spectral function in the first place without solving the quasiparticle equation.

\subsection{Inner Peaks and Spectral Functions at X2C level\label{subsec:inner-ip}}
The analytical continuation, based on the theory of Nevanlinna functions, that we employ here provides accurate information about the spectral function and ionization peaks that are energetically much deeper with respect to the chemical potential, or the Fermi level.
This allows us to elicit inner ionization peaks and resolve small spin-orbit splittings, in addition to calculating the IPs.
Here, we first consider three halides of Cadmium, CdX$_2$ (X = Cl, Br, I) as representative examples and compare the $GW$ spectral functions against experimental photoelectron spectra.
We then switch our attention to the $d$-orbital splitting for ZnCl$_2$, CdCl$_2$ and HgCl$_2$.
All of these molecules have been selected from the \texttt{SOC-81} data-set.

\subsubsection{Valence IP Spectra for CdX$_2$}\label{sec:cdx2}
% Here we look at the spectra of CdX$_2$ (X = Cl,Br,I) where experimental PES curves are available.\cite{BOGGESS1973}
% As we move from Cl to I, the SOC effect is larger.
Cadmium Halides have a linear structure with $\Pi_g$ as the highest occupied orbital, in which the valence orbitals from the halogen is excited in the first ionization.
The ordering of valence orbitals in the absence of spin-orbit coupling is ${}^2\Pi_g > {}^2\Pi_u > {}^2\Sigma_u >  {}^2\Sigma_g$.
In the presence of spin-orbit coupling, the doubly degenerate ${}^2\Pi_g$ states split into ${}^2\Pi_{\frac{3}{2}g}$, ${}^2\Pi_{\frac{1}{2}g}$  states and the doubly degenerate ${}^2\Pi_u$ splits into ${}^2\Pi_{\frac{3}{2}u}$, ${}^2\Pi_{\frac{1}{2}u}$ states.
The splitting is more significant as we go from Cl to I and should be seen in the spectra.

In ~\cref{fig:cdx}, we compare the experimental\cite{BOGGESS1973} and sc$GW$ spectral functions for CdX$_2$ (X = Cl, Br, I).
First of all, we notice that scalar relativistic (sfX2C) theory predicts an almost identical shape for the spectrum of all the halides.
This, however, is not the case for experiment.
On the other hand, by accounting for the spin-orbit splitting, the two-component approach (X2C) produces spectra which are in excellent agreement with the experiment.
%The magnitude of the spin-orbit coupling increases from Cl to I.
For Cl, we can see that the two-component relativistic calculation and the scalar relativistic corrections produces qualitatively similar results.
For Br and I, the SOC effects are quite visible and one has to rely on the X2C results to obtain accurate photoelectron spectra.
%For both CdBr$_2$ and CdI$_2$, for the $\sigma$-type MO's, it can be seen that the ${}^2\Sigma_u$ state has larger spin-orbit splitting compared to ${}^2\Sigma_g$.
%\textcolor{magenta}{(GH: I do not see any splitting in the $\Sigma$  orbitals.)}

%\subsubsection{Issue with Expermental reference and resolution of spectra}
% One of the advantages of such an accurate resolution in the theoretical spectrum is that it can provides detailed explanation and insights which is useful characterizing the different peaks in an experimental spectrum.
% For instance, when SOC is small but non-zero (e.g., in CdCl$_2$), two different peaks cannot be resolved in experiments. However, by controlling the peak-broadening (or $\eta$) in our analytical continuation, we can get results at much higher precision.

\begin{figure}
    \centering
    \includegraphics[width=\linewidth]{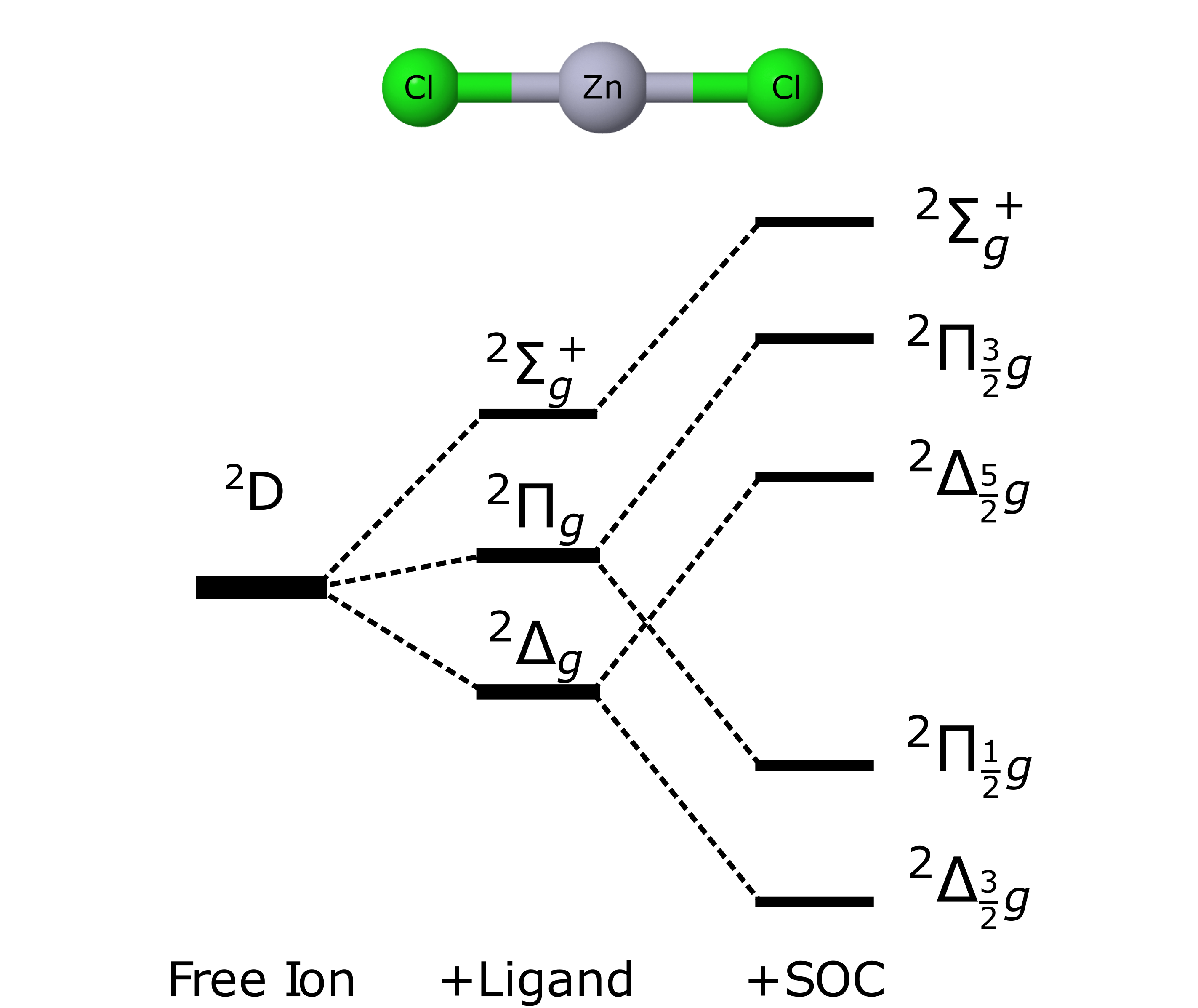}
    \caption{Pictorical depiction of splitting of the states of $d$ orbital in a linear geometry. The ordering of the %$^2\Sigma_g^{+}$ , $^2\Pi_g$ and $^2\Delta_g$ 
    states depends on the metal and the ligands. Here we show the ordering for ZnCl$_2$.}
    \label{fig:d_splitting}
\end{figure}

\begin{figure*}
    \centering
    \includegraphics[width=\linewidth]{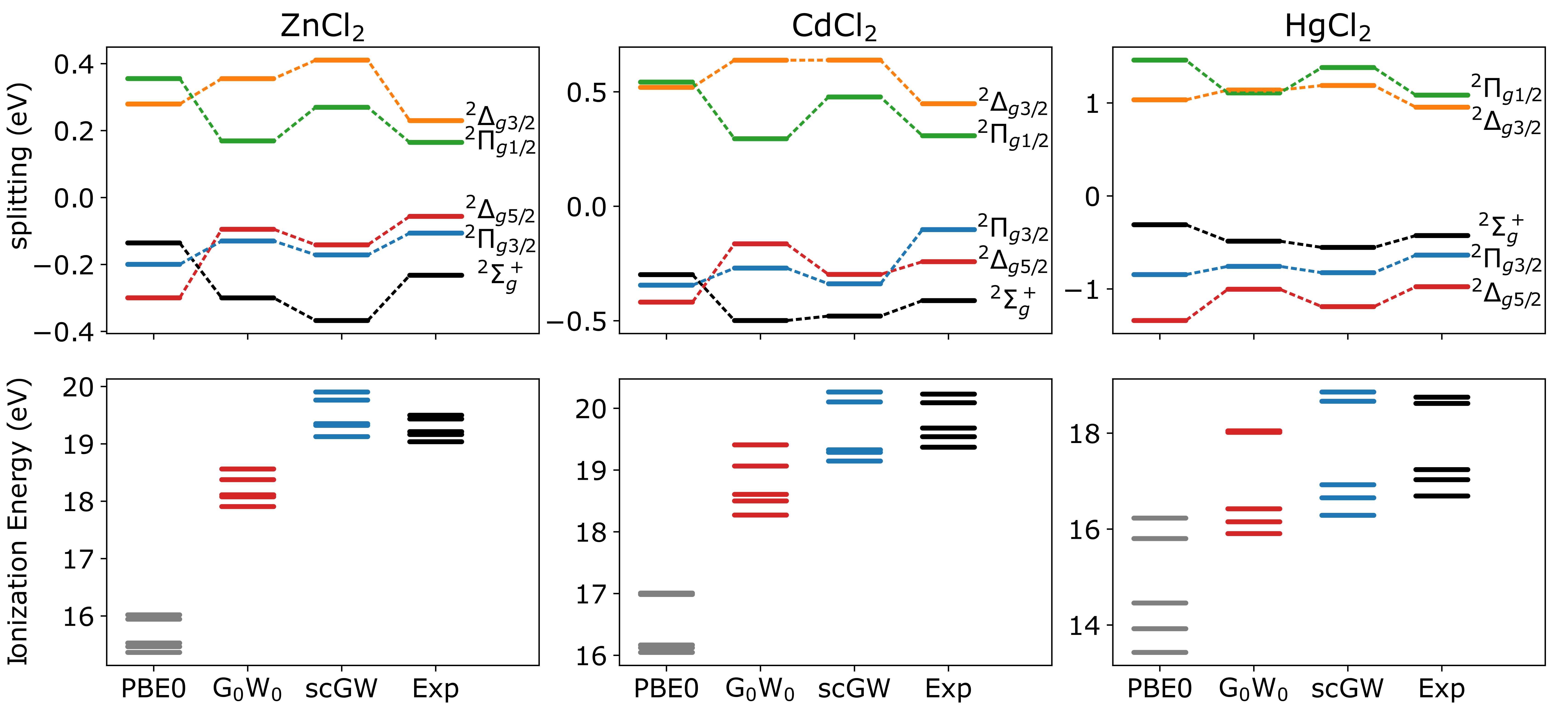}
    \caption{Ionization energy for the $d^{10} \xrightarrow{} d^{9}$ excitation for ZnCl$_2$, CdCl$_2$ and HgCl$_2$. 
    In the lower panel, we show the absolute IP's from PBE0, $\gowo$@PBE0 and sc$GW$ level compared with experiment.\cite{Lee1981,Wang1990} 
    In the top panel, we present the relative splitting of the states due to spin-orbit coupling and ligands.
    The peak labels and relative ordering are shown in the top panel. The $\gowo$ calculation is using the PBE0 reference.
    }
    \label{fig:d_absolute}
\end{figure*}
%\subsubsection{$d$ orbital splitting for Group IIB Chlorides: Zn, Cd and Hg}
\subsubsection{$d$-orbital splitting for Zn, Cd and Hg chlorides}
The photoelectron spectra of Zn, Cd and Hg compounds also exhibit inner lying peaks that arise from outer-shell $d$-orbital excitations.
These systems have $d^{10}$ configuration in the neutral state.
Hence, the resultant peaks correspond to a $d^{10} \xrightarrow{}d^{9}$ excitation.
In the absence of spin-orbit coupling and ligands, all the $d$ orbitals for a free ion are degenerate.
The interaction with a chlorine (Cl) atom induces a splitting of the inner $d$ orbitals into three states, as described by crystal field theory. \cite{cft1964} 
The $^2\Sigma_g^+$ state corresponds to the $d_{z^2}$ orbital oriented along the bond direction, while $^2\Pi_g$ ($d_{zx}$, $d_{zy}$) exhibits one node along the inter-nuclear axis.
Additionally, $^2\Delta_g$ ($d_{x^2-y^2}$, $d_{xy}$) possesses two nodes along the inter-nuclear axis.
Incorporating spin-orbit coupling at the X2C level further splits the $^2\Pi_g$ state into $^2\Pi_{g3/2}$ and $^2\Pi_{g1/2}$. Similarly, the delta-type orbitals ($^2\Delta_g$) split into $^2\Delta_{g5/2}$ and $^2\Delta_{g3/2}$ states.
Consequently, we get a total of five non-degenerate states, as illustrated in \cref{fig:d_splitting}.

The ordering of these peaks depend on two key factors: the electrostatic interaction between the ligand and metal, determining the sequence of \{$\Sigma$, $\Pi$, and $\Delta$\}, and the spin-orbit interaction, responsible for splitting the $\Delta$ and $\Pi$ states. 
The experimental findings indicate that the spin-orbit coupling is larger than the ligand field splitting.\cite{ELAND1970,Wang1990}
%It is experimentally determined that the spin-orbit coupling predominates over the splitting between the $\Delta$ and $\Pi$ states.

%In most of the older experiments,\cite{ELAND1970,berkowitz1974,Lee1981,Cocksey1973,ORCHARD1975,Furlani} only three peaks are visible: one corresponding to $^2\Sigma_g^+$, another one corresponding to $^2\Pi_{g3/2}$ and $^2\Delta_{g5/2}$, and the third corresponding to the $^2\Pi_{g1/2}$ and $^2\Delta_{g3/2}$ states.
%This is because the $\Delta_g$ and $\Pi_g$ are very close in energy.
%Lee and Potts used intensity difference between He I and He II ionization peaks and a ligand field semi-empirical model to assign the peaks for ZnCl$_2$, CdCl$_2$ and HgCl$_2$.~\cite{Lee1981}
%In their work, they assumed the $\Delta_g$ peaks to have larger intensity and recommended an ordering.
%Later, Wang et al. produced well resolved experimental peaks for ZnCl$_2$ excitations.\cite{Wang1990}
%Here we use the reference values and ordering recommended by these two experimental references.

In older experiments,\cite{ELAND1970,berkowitz1974,Lee1981,Cocksey1973,ORCHARD1975,Furlani} only three distinct peaks were observed, aligning with
(i) $^2\Sigma_g^+$, 
(ii) $^2\Pi_{g3/2}$ and $^2\Delta_{g5/2}$, and
(iii) $^2\Pi_{g1/2}$ and $^2\Delta_{g3/2}$ states.
This is due to the close proximity between $\Delta_g$ and $\Pi_g$ orbital energies.
One such experiment by Lee and Potts used intensity variation between He I and He II ionization peaks, in combination with a ligand-field semi-empirical model to characterize ZnCl$_2$, CdCl$_2$, and HgCl$_2$.~\cite{Lee1981}
Wang et al. later provided resolved experimental peaks for ZnCl$_2$ excitations.~\cite{Wang1990}
We rely on the reference values and ordering recommended by both of these experimental works.
On the theoretical front, characterizing the peak structures of ZnCl$_2$ has produced contradicting results.\cite{Cederbaum1986,BAUSCHLICHER1990}

%
%The 2ph-TDA~\cite{Cederbaum1986} (Tamm-Dancoff) results by XA, and MR-CI~\cite{BAUSCHLICHER1990} results by XB have produced different orderings of the ionization states.
%Neither of these works included SOC splitting, and their results focused just on assigning the $\Pi$, $\Delta$ and $\Sigma$ characters to the ionization peaks.
%Experimentally, the peak labels and values are then determined by relative intensity between PES spectra of different resolutions and with help of semi-empirical models.\cite{Lee1981}

% We compare the experimental values for the ionization of the $d^{10} \xrightarrow{} d^{9}$ orbitals against our $GW$ results.
In the lower panel of \cref{fig:d_absolute}, we compare the experimental values of IP for the outermost, fully-filled $d$-orbitals against PBE0, $\gowo$@PBE0 and sc$GW$.
For all systems, we can see that the PBE0 quasiparticle excitation energies  are off from the experiment by more than 2 eV,
while $\gowo$@PBE0 improve these trends slightly.
The $\gowo$@PBE0 energies for $d$ orbitals are still lower by ~ 1 eV. 
This behaviour is also seen in solids for $\gowo$.\cite{Gant2022}
Even though none of the methods fully match with experiment, the fully self-consistent $GW$ provides the best agreement for absolute ionization energies.

We also compare the ordering and relative separation of the peaks in the upper panel of \cref{fig:d_absolute} for the same systems.
The experimental peak ordering is presented and it changes as we go from Zn to Hg.
Zn and Cd are more electropositive and form ionic bonds with the Cl $p$-orbitals, leaving the $d$-orbitals primarily non-bonded. Therefore, the energetic order of the orbitals follow directly from the crystal field theory.
On the other hand, Hg forms more covalent bond due to its low electropositivity.
In this case, the $d^{10}$-orbitals are more involved in the formation of bonds and $^2\Sigma_g^+$ and $^2\Pi_g$ become more stabilized compared to the $\Delta_g$ state.
Therefore, the ligand-field splitting ordering is reversed in the case of HgCl$_2$.
At the PBE0 level, all the $d^{10} \xrightarrow{} d^{9}$ ionization peaks for Zn and Cd are ordered incorrectly.
For ZnCl$_2$, this order gets corrected by both sc$GW$ and $\gowo$@PBE0.
For CdCl$_2$, however, the both $GW$ methods predict incorrect order for the $^2\Delta_{g5/2}$ and $^2\Pi_{g3/2}$ orbitals compared to experiment. 
For HgCl$_2$, the ordering is correct already at the PBE0 level but the splitting is overestimated.
There is a small ordering swap for $\gowo$ for $^2\Delta_{g3/2}$ and $^2\Pi_{g1/2}$ states as seen in the top panel of \cref{fig:d_absolute},
which gets corrected at the sc$GW$ level.

Both $\gowo$ and sc$GW$ methods provide good absolute peak location for the inner $d$ orbital excitations.
The SOC splitting of the $^2\Delta_{g}$ and $^2\Pi_{g}$ states is overestimated in all three systems within the sc$GW$ method, possibly due to the ``1e'' approximation in X2C. 
Including relativistic corrections in the two-electron interactions is known to decrease these splittings, providing more accurate results.
Computation of the spectra for these states using methods like $\Delta$SCF can be tricky since one has to optimize all 5 different excited states very close in energy.
Another alternative is to use wavefunction based methods where the computational cost can be quite high.

\subsection{Total Energies: Equilibrium bond length and harmonic frequencies\label{subsec:total-energy}}

Another advantage of sc$GW$ is that thermodynamic quantities, such as the total energy, are well defined unlike $\gowo$.
This allows access to properties directly derivable from energy trends.
In this section, using sc$GW$ energies, we calculate two such properties: equilibrium bond lengths and harmonic frequencies, for selected diatomic Halogen molecules, and compare them against reference experimental data. We also show Hartree-Fock and PBE0 estimates for comparison.

The calculations were carried out using the \texttt{X2C-QZVPall} basis set. To determine the bond lengths and harmonic frequencies of each diatomic molecule, computations were performed at 40 equidistant points, with intervals of 0.005 $\mathrm{\AA}$ along the potential energy curve surrounding the equilibrium structure. Subsequently, the obtained energy values were utilized to fit an eighth-order polynomial curve. From the coefficients of this polynomial, we derive the equilibrium bond lengths and corresponding harmonic frequencies.
Experimental data is obtained from NIST WebBook.\cite{NIST}

% Please add the following required packages to your document preamble:
% \usepackage{multirow}
\begin{table*}[]
\caption{Equilibrium bond length (in $\mathrm{\AA}$) and harmonic frequencies (in cm$^{-1}$) for the Halogen dimers at sfX2C and X2C level. We compare the results from HF, PBE0 and sc$GW$. Experimental results are reported from NIST.\cite{NIST}% or from Lans paper\cite{Liu2018}  huber paper.  
}
\renewcommand{\arraystretch}{1.3}
\begin{tabular}{llrrrlrrrlrlrrrlrrrlr}
\hline
\hline
\multicolumn{1}{c}{\multirow{3}{*}{system}} &  & \multicolumn{7}{c}{bond length ($\mathrm{\AA}$)}                                                                                                                            &  & \multicolumn{1}{c}{}           &  & \multicolumn{9}{c}{Harmonic Frequency (cm$^{-1}$)}    \\ \hline

\multicolumn{1}{c}{}                        &  & \multicolumn{3}{c}{sfX2C}                                                    &  & \multicolumn{3}{c}{X2C}                                                      &  & \multicolumn{1}{l}{Experiment} &  & \multicolumn{3}{c}{sfX2C}                                                    &  & \multicolumn{3}{c}{X2C}                                                      &  & \multicolumn{1}{l}{Experiment} \\ 
                                            &  & \multicolumn{1}{c}{HF} & \multicolumn{1}{c}{PBE0} & \multicolumn{1}{c}{sc$GW$} &  & \multicolumn{1}{c}{HF} & \multicolumn{1}{c}{PBE0} & \multicolumn{1}{c}{sc$GW$} &  & \multicolumn{1}{l}{}           &  & \multicolumn{1}{c}{HF} & \multicolumn{1}{c}{PBE0} & \multicolumn{1}{c}{sc$GW$} &  & \multicolumn{1}{c}{HF} & \multicolumn{1}{c}{PBE0} & \multicolumn{1}{c}{sc$GW$} &  & \multicolumn{1}{l}{}           \\
\hline
                                            
Cl$_2$                                         &  & 1.975                  & 1.981                    & 1.977                    &  & 1.975                  & 1.981                    & 1.977                    &  & \textbf{1.988}                 &  & 577.6                  & 543.4                    & 551.9                    &  & 577.5                  & 543.2                    & 552.0                    &  & \textbf{560}                   \\
Br$_2$                                         &  & 2.271                  & 2.277                    & 2.271                    &  & 2.273                  & 2.281                    & 2.275                    &  & \textbf{2.281}                 &  & 333.2                  & 316.0                    & 323.3                    &  & 330.7                  & 312.6                    & 319.6                    &  & \textbf{325}                   \\
I$_2$                                          &  & 2.661                  & 2.655                    & 2.644                    &  & 2.673                  & 2.671                    & 2.657                    &  & \textbf{2.665}                 &  & 223.7                  & 215.5                    & 217.5                    &  & 215.8                  & 205.4                    & 214.2                    &  & \textbf{215}                   \\
At$_2$                                         &  & 2.839                  & 2.831                    & 2.817                    &  & 2.920                  & 2.924                    & 2.885                    &  &   
&  & 159.9                  & 154.8                    & 160.1                    &  & 132.7                  & 124.5                    & 135.4                    &  &       
\\ \hline \hline
\end{tabular}
\label{tbl:energy}
\end{table*}

\begin{figure}
    \centering
    \includegraphics[width=\linewidth]{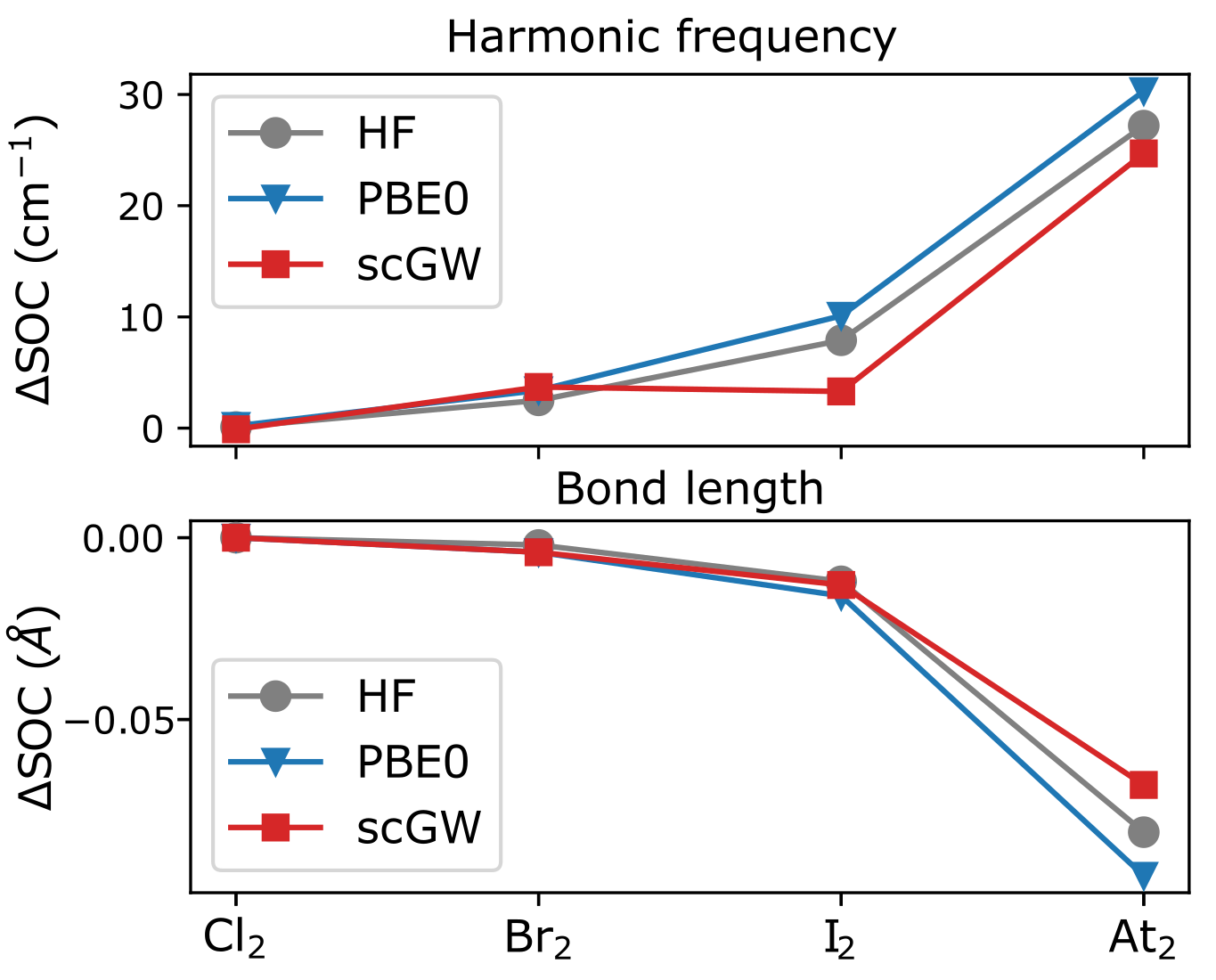}
    \caption{Relativistic correction due to spin-orbit coupling on the equilibrium bond length and harmonic frequencies of the halides at HF, PBE0 and sc$GW$ level. }
    \label{fig:deltasoc}
\end{figure}

We report both sfX2C and X2C results in  \cref{tbl:energy}.
For bond length, we can see that for Cl$_2$, Br$_2$ and I$_2$ there is little to no effect of SOC correction on top of the sfX2C reference at HF and sc$GW$ level.
For At$_2$, the bond length does change significantly but still in second decimal place.
On the other hand, as we go from sfX2C to X2C, the inclusion of SOC effects reduces the harmonic frequencies.
HF seems to overestimate the frequencies while sc$GW$ underestimates them, at least for these diatomic systems studied.

Overall, at the X2C level, sc$GW$ demonstrates agreement with experimental bond lengths within a margin of 0.01 $\mathrm{\AA}$, and harmonic frequencies within 10 cm$^{-1}$. 
While PBE0 yields highly accurate bond lengths, it shows slightly less precision in harmonic frequencies compared to the sc$GW$.
In \cref{fig:deltasoc} we present the shift in bond length and harmonic frequencies as we go from sfX2C to X2C approximation. This highlights the magnitude of corrections due to SOC, and shows an expected increasing trend with respect to the size of atom.

For very accurate results to within 1 cm$^{-1}$, one needs to use very large basis sets and sometimes also have to un-contract them.
%For systems as large as At, a general basis set is also not available which is very well tested and benchmarked.
Poor basis-set convergence and a lack of accurate Breit term in the Hamiltonian make the computation of accurate harmonic frequencies a challenging task.\cite{Tew2007,Vasiliu2021,Liu2018,Zhang2022}

\section{Conclusion}
In this study, we have presented a comprehensive analysis of the performance of the fully self-consistent $GW$ with both scalar (sfX2C) and two-component (X2C) relativistic effects in Gaussian basis sets for molecules.
Our results demonstrate that the fully self-consistent $GW$ method produces accurate IPs and energies for heavy elements.
Among the one-shot $GW$ variants, we see heavy dependence of the DFT reference.
We note that $\gowo$ with a hybrid DFT reference (PBE0) also gives good estimation of IP compared to experiment.

Using Nevanlinna analytic continuation, we analyze the spectral function for few of the systems in the \texttt{SOC-81} molecular dataset, and find excellent agreement between sc$GW$ and experimental photoelectron spectra.
% We see excellent agreement between analytically continued sc$GW$ spectra and experimental photoelectron spectra.
Inclusion of SOC effects become important for comparison with experimental spectra as we go to heavier elements.
We also see excellent agreement for inner $d$-shell excitation for selected representative examples.
This work is also the first step towards calculation of harmonic frequencies and bond lengths with heavy elements using sc$GW$ and further study is required to converge the basis, relativistic corrections to access the accuracy.

The low computational cost of sc$GW$ makes it a good alternative to more expensive wavefunction based methods, especially for systems with heavy element.
sc$GW$ is also a good alternative to $\gowo$ since it removes the dependence of the DFT reference with a pre-factor in computational cost.
For heavy element systems, to include the relativistic effects at the core and valence level, very large, and sometime uncontracted, basis-sets are required and is still a challenge in obtaining very accurate results in correlated calculations.\cite{Cheng2017}
To improve on this, one of the future directions is to implement an RPA frozen natural orbital (FNO) based relativistic self-consistent $GW$.\cite{Surjuse2022,Yuan2022,Ramberger2019}
In future, we will also include two-electron contribution to the spin-orbit effects at mean-field level for accurate inner core and heavy element calculations.\cite{Dmitry2013,Cheng2017}

\begin{acknowledgements}
V. A., G. H. and D. Z. are supported by the U.S. Department of Energy, Office of Science, Office of Advanced Scientific Computing
    Research and Office of Basic Energy Sciences, Scientific
    Discovery through Advanced Computing (SciDAC) program under Award Number DE-SC0022198.
This research used resources of the National Energy Research Scientific Computing Center,
    a DOE Office of Science User Facility supported by the Office of Science of the U.S. Department of Energy
    under Contract No. DE-AC02-05CH11231 using NERSC award BES-ERCAP0027729.
We gratefully acknowledge helpful discussions with Dr. Chia-Nan Yeh and Dr. Pavel Pokhilko.
\end{acknowledgements}

\section{Supporting Information} 
Ionization potential for \texttt{SOC-81} at sfX2C and X2C level for sc$GW$, $\gowo$@PBE and $\gowo$@PBE0. 
Discussion on comparison of the Gaussian based $GW$ with plane-wave based $GW$ and slater-type orbital $GW$. 
Discussion on the multiple solutions of the QP equation for SrO.

\bibliographystyle{achemso}
\bibliography{library}
\end{document}

\title{Relativistic fully self consistent GW for molecules: Total energies and ionization potentials
  } 

\author{Vibin Abraham}
%\email{avibin@umich.edu}

\author{Gaurav Harsha}
\affiliation{%
 Department of Chemistry, University of Michigan, Ann Arbor, MI 48109.
}%
\author{Dominika Zgid}
%\email{zgid@umich.edu}
\affiliation{%
 Department of Chemistry, University of Michigan, Ann Arbor, MI 48109.
}%
\affiliation{%
 Department of Physics and Astronomy, University of Michigan, Ann Arbor, MI 48105.
}%

\maketitle
\setlength\LTcapwidth{\linewidth}
\newcolumntype{L}[1]{>{\raggedleft\arraybackslash}p{\dimexpr#1\textwidth-2\tabcolsep\relax}}
\newcolumntype{R}[1]{>{\raggedright\arraybackslash}p{\dimexpr#1\textwidth-2\tabcolsep\relax}}

\section{Data for SOC-81}
In this section we present our data for the SOC-81 database using the three levels of theory, $G_0W_0$@PBE, $G_0W_0$@PBE0 and sc$GW$.
We present the data at the scalar relativisitic and the two-component level.
The data is shown in Table \ref{tbl:full_data}.

\renewcommand*{\arraystretch}{0.7}
\begin{longtable}{R{0.15}L{0.11}L{0.11}L{0.11}L{0.11}L{0.01}L{0.11}L{0.11}L{0.11}}% Adjust the width as needed
\caption{Comparison of experimental data against $G_0W_0$ and sc$GW$ results for ionization potentials (IP) for the \texttt{SOC-81} dataset at both the sfX2C and X2C levels of theory. Calculated values are extrapolated using the inverse of basis functions\cite{van2015, forster2023}. * denotes experimental references with corrections for KBr and RbBr from \cite{GOODMAN1974} and for HgCl$_2$ from \cite{ELAND1970}.} \label{tbl:full_data} \\

\toprule[1pt] 

\hline \hline
        & \multicolumn{1}{l}{}     & \multicolumn{3}{c}{sfX2C}                                                                   &  & \multicolumn{3}{c}{X2C}                                                                     \\ \cline{3-5} \cline{7-9} 
System  & \multicolumn{1}{c}{Experiment} & \multicolumn{1}{c}{$G_0W_0$@PBE} & \multicolumn{1}{c}{$G_0W_0$@PBE0} & \multicolumn{1}{c}{sc$GW$} &  & \multicolumn{1}{c}{$G_0W_0$@PBE} & \multicolumn{1}{c}{$G_0W_0$@PBE0} & \multicolumn{1}{c}{sc$GW$} \\ \hline 
\hline 

\endfirsthead
\hline
\hline%
        & \multicolumn{1}{l}{}     & \multicolumn{3}{c}{sfX2C}                                                                   &  & \multicolumn{3}{c}{X2C}                                                                     \\ \cline{3-5} \cline{7-9} 
System  & \multicolumn{1}{c}{Experiment} & \multicolumn{1}{c}{$G_0W_0$@PBE} & \multicolumn{1}{c}{$G_0W_0$@PBE0} & \multicolumn{1}{c}{sc$GW$} &  & \multicolumn{1}{c}{$G_0W_0$@PBE} & \multicolumn{1}{c}{$G_0W_0$@PBE0} & \multicolumn{1}{c}{sc$GW$} \\ \hline 
\hline
\endhead
\hline
\endfoot
\endlastfoot
  1 AgBr               & 9.59                     & 9.70                          & 9.73                           & 9.53                       &  & 9.57                          & 9.60                           & 9.38                       \\
  2 AgCl               & 10.08                    & 10.16                         & 10.23                          & 10.09                      &  & 10.18                         & 10.19                          & 10.07                      \\
  3 AgI                & 8.80                     & 9.33                          & 9.22                           & 8.86                       &  & 9.11                          & 8.98                           & 8.56                       \\
  4 AlBr$_3$           & 10.91                    & 10.79                         & 11.10                          & 11.00                      &  & 10.74                         & 11.04                          & 10.88                      \\
  5 AlI$_3$            & 9.66                     & 9.68                          & 9.94                           & 9.66                       &  & 9.51                          & 9.75                           & 9.46                       \\
  6 AsBr$_3$           & 10.21                    & 10.17                         & 10.35                          & 10.04                      &  & 10.12                         & 10.30                          & 10.00                      \\
  7 AsCl$_3$           & 10.90                    & 10.80                         & 11.05                          & 10.68                      &  & 10.80                         & 11.05                          & 10.71                      \\
  8 AsF$_3$            & 13.00                    & 12.73                         & 13.02                          & 12.94                      &  & 12.73                         & 13.02                          & 12.92                      \\
  9 AsH$_3$            & 10.58                    & 10.37                         & 10.51                          & 10.30                      &  & 10.36                         & 10.51                          & 10.30                      \\
 10 AsI$_3$            & 9.00                     & 9.44                          & 9.49                           & 9.79                       &  & 9.21                          & 9.32                           & 8.95                       \\
 11 Br$_2$             & 10.54                    & 10.85                         & 10.83                          & 10.56                      &  & 10.74                         & 10.68                          & 10.36                      \\
 12 BrCl               & 11.01                    & 11.21                         & 11.20                          & 10.94                      &  & 11.11                         & 11.09                          & 10.80                      \\
 13 C$_2$H$_2$Se       & 8.71                     & 8.91                          & 8.82                           & 8.43                       &  & 8.90                          & 8.81                           & 8.41                       \\
 14 C$_2$HBrO          & 9.10                     & 9.42                          & 9.48                           & 9.15                       &  & 9.43                          & 9.47                           & 9.12                       \\
 15 C$_4$H$_4$Se       & 8.86                     & 8.73                          & 9.00                           & 8.56                       &  & 8.74                          & 9.00                           & 8.59                       \\
 16 CF$_3$I            & 10.45                    & 10.73                         & 10.95                          & 10.64                      &  & 10.48                         & 10.67                          & 10.30                      \\
 17 C$_2$H$_6$Cd          & 8.80                     & 8.91                          & 9.11                           & 8.86                       &  & 8.90                          & 9.11                           & 8.88                       \\
 18 C$_2$H$_6$Hg          & 9.32                     & 9.11                          & 9.25                           & 8.96                       &  & 9.11                          & 9.26                           & 8.99                       \\
 19 C$_2$H$_6$Se          & 8.40                     & 8.25                          & 8.39                           & 8.19                       &  & 8.24                          & 8.39                           & 8.18                       \\
 20 C$_2$H$_6$Zn          & 9.40                     & 9.48                          & 9.67                           & 9.45                       &  & 9.48                          & 9.68                           & 9.46                       \\
 21 CH$_3$HgBr         & 10.16                    & 9.86                          & 10.15                          & 10.07                      &  & 9.74                          & 10.00                          & 9.86                       \\
 22 CH$_3$HgCl         & 10.84                    & 10.39                         & 10.74                          & 10.68                      &  & 10.35                         & 10.69                          & 10.72                      \\
 23 CH$_3$HgI          & 9.25                     & 9.39                          & 9.46                           & 9.23                       &  & 9.18                          & 9.20                           & 8.89                       \\
 24 CH$_3$I            & 9.52                     & 9.66                          & 9.80                           & 9.53                       &  & 9.43                          & 9.54                           & 9.21                       \\
 25 CI$_4$             & 9.10                     & 9.39                          & 9.50                           & 9.16                       &  & 9.17                          & 9.27                           & 9.10                       \\
 26 CaBr$_2$           & 10.35                    & 10.02                         & 10.33                          & 10.33                      &  & 9.93                          & 10.21                          & 10.16                      \\
 27 CaI$_2$            & 9.39                     & 9.40                          & 9.62                           & 9.44                       &  & 9.21                          & 9.38                           & 9.21                       \\
 28 CdBr$_2$           & 10.58                    & 10.47                         & 10.75                          & 10.64                      &  & 10.35                         & 10.62                          & 10.46                      \\
 29 CdCl$_2$           & 11.44                    & 11.11                         & 11.46                          & 11.32                      &  & 11.08                         & 11.42                          & 11.32                      \\
 30 CdI$_2$            & 9.57                     & 9.69                          & 9.90                           & 9.69                       &  & 9.45                          & 9.63                           & 9.30                       \\
 31 CsCl               & 8.69                     & 8.29                          & 8.63                           & 8.54                       &  & 8.26                          & 8.60                           & 8.50                       \\
 32 CsF                & 9.68                     & 8.73                          & 9.37                           & 9.81                       &  & 8.74                          & 9.38                           & 9.86                       \\
 33 CsI                & 7.46                     & 7.61                          & 7.73                           & 7.45                       &  & 7.44                          & 7.51                           & 7.18                       \\
 34 CuF                & 10.90                    & 9.98                          & 10.38                          & 10.71                      &  & 10.55                         & 10.35                          & 10.67                      \\
 35 HgCl$_2$*           & \textbf{11.37}                    & 11.05                         & 11.42                          & 11.23                      &  & 10.96                         & 11.34                          & 11.18                      \\
 36 I$_2$              & 9.35                     & 9.86                          & 9.79                           & 9.40                       &  & 9.55                          & 9.50                           & 9.02                       \\
 37 IBr                & 9.85                     & 10.29                         & 10.21                          & 9.88                       &  & 10.04                         & 9.96                           & 9.53                       \\
 38 ICl                & 10.10                    & 10.52                         & 10.45                          & 10.08                      &  & 10.27                         & 10.18                          & 9.79                       \\
 39 IF                 & 10.62                    & 10.97                         & 10.89                          & 10.61                      &  & 10.67                         & 10.56                          & 10.26                      \\
 40 KBr*                & \textbf{8.10}                     & 7.58                          & 8.31                           & 8.19                       &  & 7.41                          & 8.20                           & 8.12                       \\
 41 KI                 & 7.40                     & 7.60                          & 7.90                           & 7.68                       &  & 7.38                          & 7.71                           & 7.35                       \\
 42 Kr$_2$             & 13.77                    & 13.58                         & 13.82                          & 13.85                      &  & 13.49                         & 13.71                          & 13.65                      \\
 43 KrF$_2$            & 13.34                    & 13.08                         & 13.60                          & 13.90                      &  & 13.03                         & 13.54                          & 13.83                      \\
 44 LaBr$_3$           & 10.68                    & 10.15                         & 10.62                          & 10.53                      &  & 9.98                          & 10.50                          & 10.43                      \\
 45 LaCl$_3$           & 11.29                    & 10.85                         & 11.28                          & 11.17                      &  & 10.87                         & 11.29                          & 11.18                      \\
 46 LiBr               & 9.44                     & 9.24                          & 9.38                           & 9.29                       &  & 8.88                          & 9.28                           & 9.14                       \\
 47 LiI                & 8.44                     & 8.59                          & 8.78                           & 8.56                       &  & 8.44                          & 8.55                           & 8.25                       \\
 48 MgBr$_2$           & 10.85                    & 10.69                         & 11.02                          & 10.93                      &  & 10.58                         & 10.89                          & 10.86                      \\
 49 NaBr               & 8.70                     & 8.39                          & 8.82                           & 8.69                       &  & 8.29                          & 8.71                           & 8.53                       \\
 50 NaI                & 8.00                     & 8.29                          & 8.37                           & 8.08                       &  & 8.08                          & 8.14                           & 7.78                       \\
 51 OsO$_4$            & 12.35                    & 12.20                         & 12.68                          & 12.33                      &  & 12.20                         & 12.75                          & 12.36                      \\
 52 PBr$_3$            & 10.00                    & 9.92                          & 10.16                          & 9.83                       &  & 9.90                          & 10.13                          & 9.81                       \\
 53 POBr$_3$           & 11.03                    & 10.97                         & 11.35                          & 11.14                      &  & 10.88                         & 11.16                          & 11.06                      \\
 54 RbBr*               & \textbf{8.00}                     & 7.08                          & 8.27                           & 8.10                       &  & 7.19                          & 8.18                           & 8.01                       \\
 55 RbCl               & 8.70                     & 8.34                          & 8.70                           & 8.62                       &  & 8.29                          & 8.66                           & 8.58                       \\
 56 RbI                & 7.30                     & 7.73                          & 7.84                           & 7.56                       &  & 7.55                          & 7.62                           & 7.32                       \\
 57 RuO$_4$            & 12.15                    & 11.98                         & 12.48                          & 12.03                      &  & 12.34                         & 12.46                          & 12.03                      \\
 58 SOBr$_2$           & 10.59                    & 10.62                         & 10.83                          & 10.53                      &  & 10.58                         & 10.81                          & 10.46                      \\
 59 SPBr$_3$           & 9.89                     & 9.87                          & 10.00                          & 9.63                       &  & 9.81                          & 9.98                           & 9.60                       \\
 60 SeCl$_2$           & 9.50                     & 9.66                          & 9.62                           & 9.31                       &  & 9.65                          & 9.62                           & 9.30                       \\
 61 SeO$_2$            & 11.76                    & 11.66                         & 11.90                          & 11.83                      &  & 11.66                         & 11.90                          & 11.84                      \\
 62 SiBrF$_3$          & 12.46                    & 11.98                         & 12.23                          & 12.01                      &  & 11.87                         & 12.11                          & 11.84                      \\
 63 SiH$_3$I           & 9.82                     & 10.01                         & 10.20                          & 9.99                       &  & 9.78                          & 9.95                           & 9.65                       \\
 64 SrBr$_2$           & 9.82                     & 9.83                          & 10.07                          & 9.93                       &  & 9.74                          & 9.95                           & 9.79                       \\
 65 SrCl$_2$           & 10.20                    & 10.24                         & 10.56                          & 10.51                      &  & 10.23                         & 10.54                          & 10.46                      \\
 66 SrI$_2$            & 9.01                     & 9.25                          & 9.41                           & 9.23                       &  & 9.06                          & 9.17                           & 8.90                       \\
 67 SrO                & 6.91                     & 5.76                          & 6.27                           & 6.52                       &  & 5.98                          & 5.88                           & 6.54                       \\
 68 TiBr$_4$           & 10.57                    & 10.58                         & 10.84                          & 10.65                      &  & 10.48                         & 10.73                          & 10.47                      \\
 69 ZnBr$_2$           & 10.89                    & 10.73                         & 11.02                          & 10.93                      &  & 10.63                         & 10.88                          & 10.74                      \\
 70 ZnCl$_2$           & 11.80                    & 11.47                         & 11.82                          & 11.73                      &  & 11.44                         & 11.78                          & 11.71                      \\
 71 ZnF$_2$            & 13.91                    & 13.09                         & 13.68                          & 14.12                      &  & 13.06                         & 13.66                          & 14.15                      \\
 72 ZnI$_2$            & 9.73                     & 9.84                          & 10.07                          & 9.84                       &  & 9.60                          & 9.80                           & 9.50                       \\
 73 ZrBr$_4$           & 10.86                    & 10.66                         & 11.03                          & 10.87                      &  & 10.65                         & 10.92                          & 10.73                      \\
 74 ZrCl$_4$           & 11.94                    & 11.51                         & 12.01                          & 11.80                      &  & 11.50                         & 12.00                          & 11.81                      \\

\hline \hline
\bottomrule[1pt]
\end{longtable}

\section{Multiple Solutions in quasiparticle equation in $G_0W_0$}
\begin{figure}
    \centering
    \includegraphics[width=.7\linewidth]{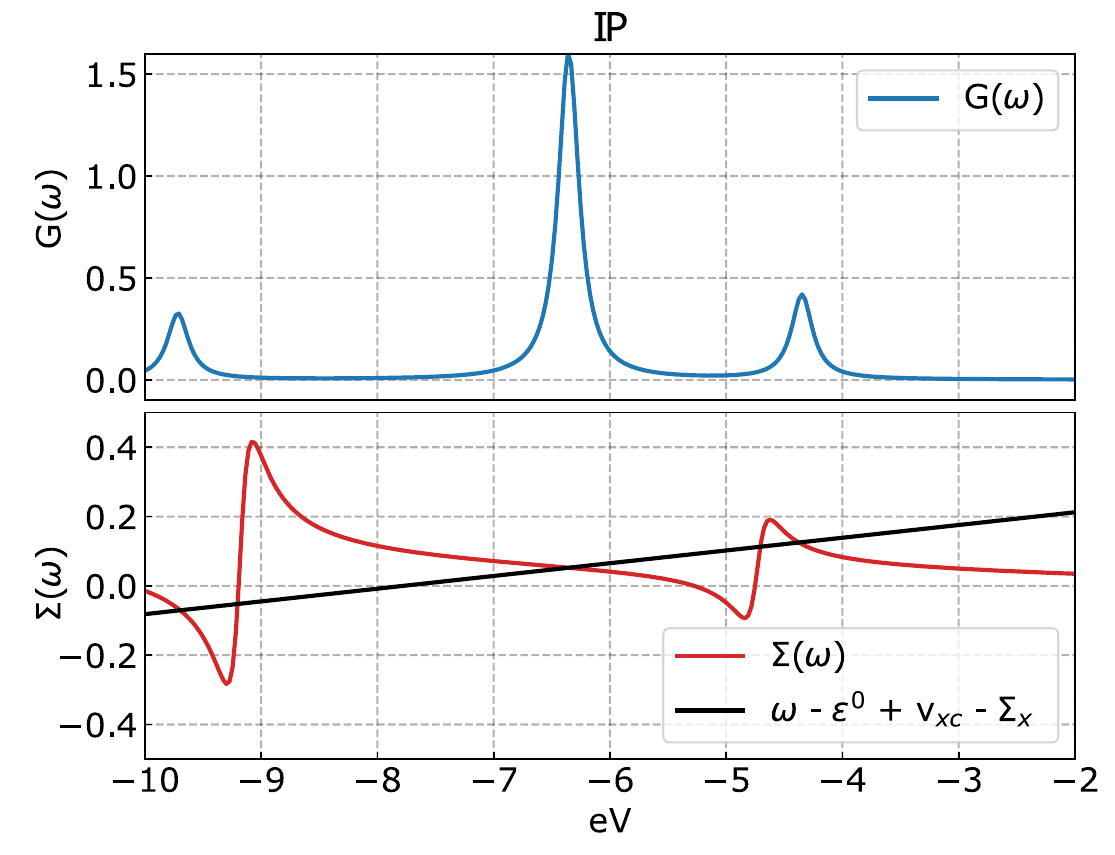}
    \caption{G$_0$W$_0$@PBE0 results for SrO:Upper Panel: G($\omega$) for SrO for HOMO. Lower Panel: Real part of correlation part of Self Energy for HOMO. }
    \label{fig:sro}
\end{figure}

The non-linearity of the quasiparticle equation implies that it can, in principle, exhibit multiple solutions.
While this does not really pose an issue generally, for some systems in the \texttt{SOC-81} testset, the iterative solution to the QP equation converges to different solutions, typically for different levels of basis set.
Graphically, solutions of the QP equation correspond to the points of intersection between $y = \omega - \epsilon_0 - \Sigma_x + v_{xc}$ and $y = \Sigma_c (\omega)$.
In Fig.~\ref{fig:sro}, we present the results for SrO for using PBE0 reference and \texttt{X2C-QZVPall} basis set.
From the bottom panel of Fig.~\ref{fig:sro}, we can say that in SrO, there are at least three possible solutions for the QP equation corresponding to HOMO, whereas from the orbital resolved spectral function in the top panel, it is clear that the highest peak in the center is the physical quasiparticle peak. 
Generally, one does not need to worry about the existence of multiple solutions because the physical solution is either unique or energetically well separated from the unphysical ones.
In case of PBE0 reference the multiple solution issue was seen only in SrO. 
For PBE reference this issue was seen in AgBr, AgCl, CuF, KBr, KI, NaBr, NaI, SeO$_2$, SrO. 

We have corrected the IP using the graphical solution for these methods for the final IP.
In the top panel of Fig.~\ref{fig:sro}, we show the quasiparticle spectrum for the G$_0$W$_0$ result for SrO.
In the bottom panel, we plot the correlation self-energy $\Sigma_c (\omega)$ as well as $y = \omega - \epsilon^{\mathrm{HOMO}}_{\mathrm{PBE0}} + v^{xc} - \Sigma^x$.
The solution of the quasiparticle equation occurs at the intersection of these two curves.

\section{Comparison with other basis sets}
The single shot GW with PBE  reference was computed by PW and STO basis earlier.
In this section, we compute the statistical quantities for PW reference\cite{Scherpelz2016} and STO\cite{forster2023} values with our GTO based values.
Due to the issue of multiple solutions to the qp equation, F\"{o}rster et al. studied a subset of 67 systems in their work due to the multiple solution issue.
Since we excluded 7 systems as mentioned in the main text, here we compare the remaining 60 systems at the PBE and PBE0 level.
We present the MAE for this data in Table \ref{tbl:basis_compare}.
As we can see, the $G_0W_0$ with the hybrid PBE0 reference gives better results compared to the PBE reference.

\begin{table}[h]    
    \caption{Comparison of mean absolute deviation of Gaussian basis $G_0W_0@PBE$ and $G_0W_0@PBE0$ with plane wave and Slater type orbital reference values in comparison to experimental results. The 1-component result is the sfX2C for us and 2-component is the X2C results.}
    \centering
    \renewcommand*{\arraystretch}{1.5}
    \begin{tabular}{m{0.15\textwidth}|m{0.1\textwidth}m{0.1\textwidth}m{0.1\textwidth}|m{0.1\textwidth}m{0.1\textwidth}m{0.1\textwidth}}
    \toprule[1pt]
    \hline \hline
    & \multicolumn{3}{c|}{$G_0W_0$@PBE} & \multicolumn{3}{c}{$G_0W_0$@PBE0} \\
    & STO & PW & GTO & STO & PW & GTO \\
    \hline
    1-component & 0.43 & 0.37 & 0.24 & 0.14 & 0.14 & 0.21 \\
    2-component & 0.52 & 0.47 & 0.23 & 0.18 & 0.13 & 0.14 \\
    \hline \hline
    \bottomrule[1pt]
    \end{tabular}
    \label{tbl:basis_compare}
\end{table}

\bibliographystyle{achemso}
\bibliography{library}